\documentstyle[12pt]{article}
\font\elevenbf=cmbx10 scaled\magstep 1
 1
 1

\begin{document}
\newcommand{\sbar}{\,\overline{\! S}}
\newcommand{\tbar}{\overline{T}}
\newcommand{\ubar}{\overline{U}}
\newcommand{\zb}{\bar{\zeta}}
\newcommand{\psibar}{\overline{\Psi}}
\newcommand{\cm}{Commun.\ Math.\ Phys.~}
\newcommand{\pr}{Phys.\ Rev.\ D~}
\newcommand{\prl}{Phys.\ Rev.\ Lett.~}
\newcommand{\pl}{Phys.\ Lett.\ B~}
\newcommand{\np}{Nucl.\ Phys.\ B~}
\newcommand{\be}{\begin{equation}}
\newcommand{\en}{\end{equation}}
\begin{titlepage}
\begin{flushright}{\tt hep-th/9510079}\\
{CERN-TH/95-257}\\
{CPTH-PC378.1095}\\[3mm]
October 1995
\end{flushright}
\vspace*{0.5cm}
\begin{center}
{\bf DUALITY SYMMETRIES IN $\bf N=2$ HETEROTIC SUPERSTRING$^\star$\\}
\vglue 5pt
\vskip 0.4in
{I. ANTONIADIS\\[1.5ex]}
{\it Ecole Polytechnique, Centre de Physique Th\'eorique,$^\dagger$
91128 Palaiseau, France}\\[2ex]
{S. FERRARA\\[1.5ex]}
{\it Theory Division, CERN, 1211 Geneva 23, Switzerland}\\[2ex]
{E. GAVA\\[1.5ex]}
{\it Instituto Nazionale di Fisica Nucleare, sez.\ di
Trieste, Italy}\\
{\it and}\\
{\it International Centre for Theoretical Physics,
I-34100 Trieste, Italy}\\[2ex]
{K.S. NARAIN\\[1.5ex]}
{\it International Centre for Theoretical Physics,
I-34100 Trieste, Italy}\\[1ex] and\\[1ex]
{T.R. TAYLOR\\[1.5ex]}
{\it Department of Physics, Northeastern University,  Boston, MA 02115,
U.S.A.\\}
\vglue 2cm
{\bf Abstract}
\end{center}
\vglue 0.3cm
{\rightskip=3pc
 \leftskip=3pc
 \noindent
We review the derivation and the basic properties of the perturbative
prepotential in $N{=}2$ compactifications of the heterotic superstring.
We discuss the structure of the perturbative monodromy group
and the embedding of rigidly supersymmetric monodromies
associated with enhanced gauge groups, at both perturbative
and non-perturbative level.
\vfill}\noindent\hrulefill\newline
{\footnotesize $^\star$Based on talks presented at SUSY-95, Palaiseau,
France, 15-19 May 1995,
at the Trieste conference on S-duality and Mirror Symmetry in String
Theory, 5-9 June 1995, at the Sixth Quantum Gravity Seminar, Moscow,
Russia, 12-19 June 1995, at the Workshop on Strings, Gravity and
Related Topics, Trieste, Italy, 29-30 June 1995, and at the
International School for Subnuclear Physics, 33rd course, Erice, Italy,
July 1995.}\newline
{\footnotesize $^\dagger$Laboratoire Propre du CNRS UPR A.0014}
\end{titlepage}
\newpage
{\elevenbf\noindent 1. Overview}
\vglue 0.2cm

Duality transformations seem to play an important role in understanding
the non-per\-tur\-ba\-tive dynamics of supersymmetric Yang-Mills theories.
Several exact results have been obtained mainly in the cases
of $N=4$ and $N=2$ extended supersymmetry due to the relation of the
BPS mass formula with the central extension of the supersymmetry
algebra \cite{centext}. Duality has been established as a
non-perturbative symmetry in the ultraviolet finite $N=4$ theories
where, in particular, their partition function has been calculated
\cite{N=4}. As a first step towards non-trivial four-dimensional
theories, $N=2$ supersymmetric Yang-Mills provide simple examples of
theories having non vanishing $\beta$-function and exhibiting interesting
properties such as asymptotic freedom. Every gauge boson is in
the same multiplet with one complex scalar and a Dirac spinor. The
vacuum of the theory is then infinitely degenerate and is characterized
by the expectation values (VEVs) of the Higgs fields which break the
non-abelian group down to its maximal abelian subgroup. The effective
two-derivative action describing the interactions of the massless
abelian vector multiplets is completely determined by its analytic
prepotential which was computed exactly \cite{N=2}. It turns out that
although perturbatively there is a non-abelian symmetry at the point
where the Higgs expectation value vanishes giving rise as usually to a
confining phase, the exact theory is always in the Higgs phase.
Moreover, there are points in the moduli space of the Higgs expectation
values where non-perturbative solitonic excitations corresponding to
monopoles and dyons become massless.

Recently, there has been considerable progress in extending these
results to supergravity in the context of superstring theory. The
additional important ingredient is the conjecture of string-string
duality according to which the heterotic and type II superstring is the
same theory at the non-perturbative level \cite{strstr}. Moreover for
$N=2$ supersymmetric compactifications the dilaton, $S$, whose
expectation value plays the role of the string coupling constant,
belongs to a hypermultiplet in type II string while in the heterotic
case it belongs to a vector multiplet. Using the fact that vector
multiplets and neutral hypermultiplets do not couple to each other in
the low energy theory, this duality provides a very powerful method for
extracting non-perturbative physics of one model from the perturbative
computations in the dual model and vice versa \cite{N=2str}.

Consider for instance the ten-dimensional heterotic $E_8\otimes E_8$
superstring compactified on the six-dimensional manifold ${\bf T^2}
\times {\bf K_3}$ with spin connection identified with the gauge
connection which gives in four dimensions an $N=2$ supersymmetry and a
gauge group $E_7\otimes E_8\otimes U(1)^{2+2}$. One factor of $U(1)^2$ is
associated to the universal dilaton vector multiplet and the
graviphoton, while the second $U(1)^2$ is associated to the
two-dimensional torus ${\bf T^2}$. At a generic point of the vector
moduli space the gauge group is broken to $U(1)^{19}$ and there are no
charged massless hypermultiplets. However there are special points where
massless charged hypermultiplets appear and their VEVs reduce the rank
of the gauge group. On the other hand in the moduli space of
hypermultiplets, there are special points where additional vector
multiplets become massless leading to an increase in the rank. As a
result one can get any rank $r$ starting from $r=2$ up to a maximum of
$2+22$ corresponding to the simple free fermionic constructions.

The classical moduli space of vector multiplets in these theories is
\be
\left. {SU(1,1)\over U(1)}\right|_{\makebox{dilaton}}\left.
\times ~{O(2,r)\over O(2)\times O(r)}\right/\Gamma
\en
where $\Gamma$ is the discrete $T$-duality group which in the simplest
case is $O(2,r;{\bf Z})$ \cite{du}. At the generic points in this moduli
space the gauge group is abelian $U(1)^r$ and there are no charged
massless states. However there are complex codimension 1 surfaces where
one of the $U(1)$'s is enhanced to $SU(2)$, due to the appearance of two
extra charged massless vector multiplets, or some charged hypermultiplets
become massless. The perturbative correction to the prepotential, which
due to the $N=2$ non-renormalization theorem occurs only at the one-loop
level, develops a logarithmic singularity near these surfaces
\cite{afgnt,dkll}. As a result the classical duality group $\Gamma$
gets modified at the perturbative level \cite{afgnt}. At the full
non-perturbative level, from the analysis in the rigid case, this
enhanced symmetry locus is expected to split into several branches where
non-perturbative states corresponding to dyonic hypermultiplets become
massless.

This class of heterotic theories is dual to the type II superstring
compactified on Calabi-Yau threefolds \cite{ceres,N=2str}. The latter is
characterized by the two Betti numbers $b_{11}$ and $b_{12}$ which
determine the number of massless vector multiplets and hypermultiplets
to be $b_{11}$ and $b_{12}+1$, respectively, where the extra $+1$ is
accounted for by the dilaton. At the perturbative level the gauge group
is abelian $U(1)^r$ with $r=b_{11}+1$ (the $+1$ accounts for the
graviphoton) and there are no charged massless matter fields. Since the
dilaton belongs to a hypermultiplet, the tree level prepotential is
exact at the full quantum level. Moreover this tree level prepotential
can be computed exactly, {\it i.e}.\ including the world-sheet instanton
corrections, by using mirror symmetry \cite{CY}. A generic feature of
the prepotential is that it has logarithmic singularities near the
conifold locus in the moduli space of the Calabi-Yau manifold
\cite{candc}. This singularity is due to the appearance of massless
hypermultiplets, corresponding to charged black holes, at the conifold
locus \cite{andy}. It can then be understood as a one-loop effect
involving this massless black hole in the internal line.

The equivalence between $N=2$ heterotic compactifications on ${\bf T^2}
\times {\bf K_3}$ and type II on Calabi-Yau has been checked explicitly
in several examples involving rank 3 and 4 \cite{cheks,monodr,w2g}. In
particular it has been shown that after identifying the heterotic
dilaton with a particular $T$-modulus of type II, the prepotential in
the type II theory reproduces the perturbative prepotential of the
corresponding heterotic model in the weak coupling limit
$S\rightarrow\infty$. Moreover at finite values of $S$, the conifold
singularity structure reproduces the exact results of the rigid
supersymmetric Yang-Mills theory \cite{monodr}. Finally, besides the
comparison of the low energy theory, further non-trivial tests have been
performed by analyzing the structure of higher dimensional interactions
\cite{w2g}.

In this talk, we review the derivation and the basic properties of the
perturbative prepotential in $N=2$ compactifications of the
heterotic superstring. We concentrate on the rank-four example which
involves, besides the dilaton, two complex moduli, $T$ and $U$,
parameterizing the two-dimensional torus ${\bf T^2}$. The associated
$U(1)\otimes U(1)$ gauge group becomes enhanced to $SU(2)\otimes U(1)$
along the $T=U$ line, and further enhanced to $SO(4)$ or to $SU(3)$ at
$T=U=i$ and $T=U=\rho({=}e^{2\pi i/3})$, respectively. At these
particular surfaces the one-loop prepotential develops logarithmic
singularities. We study the corresponding monodromies and exhibit the
resulting modifications to the classical duality transformations. The
perturbative monodromies of the rigid supersymmetric theories at the
enhanced symmetric points with the maximum non-abelian gauge symmetry
form an infinite dimensional subgroup of the full string (perturbative)
monodromy group.
\vglue 0.4 cm
{\elevenbf\noindent 2. String computation of the prepotential}
\vglue 0.2 cm

The simplest way to determine the prepotential is to reconstruct it from
the K\"ahler metric of moduli fields. Indeed, the K\"ahler potential of
the effective $N{=}2$ locally supersymmetric theory can be written as
\begin{equation}
K=-\ln (iY)\: ,~~~~~~~~~~Y=2F-2\bar{F}-{\sum_{Z=S,T,U}}
(Z-\bar{Z})(F_Z+\bar{F}_Z)\: ,
\label{Y}
\end{equation}
where $F$ is the analytic prepotential and $F_Z\equiv\partial_Z F$. Its
general form is:
\begin{equation}
F~=~STU+f(T,U)\, ,
\label{F}
\end{equation}
where the first term proportional to the dilaton, is the tree-level
contribution, and the one-loop correction is contained in a
dilaton-independent analytic function $f(T,U)$. In our conventions $S$
is defined such that $\langle S\rangle ={\theta\over\pi}+i{8\pi\over
g^2}$ where $g$ is the string coupling constant and $\theta$ the usual
$\theta$-angle. Higher loop corrections are forbidden from analyticity
and the axionic shift,
\be
D:\quad S\rightarrow S+\lambda \ ,
\label{axionshift}
\en
which is an exact continuous symmetry in string perturbation theory.

The one loop moduli metric can be obtained by expanding (\ref{Y}),
\begin{equation}
K_{Z\bar{Z}}~=~K^{(0)}_{Z\bar{Z}}[1+\frac{2i}{S-\sbar}{\cal I}+\cdots]
\quad Z=T,U\ ,
\label{metric}
\end{equation}
where the tree level metric $K^{(0)}_{Z\bar{Z}}=-(Z-\bar{Z})^{-2}$ and
$\cal I$ is given by
\begin{equation}
{\cal I}~=~-\frac{i}{2}(\partial_T-\frac{2}{T-\tbar})
(\partial_U-\frac{2}{U-\ubar})f ~+~\makebox{c.c.}
\label{Idef}
\end{equation}
${\cal I}$ can be computed by a one loop string calculation of an
amplitude involving the antisymmetric tensor using the method of
ref.\cite{yuka}. {}From the expression (\ref{Idef}) one can easily
deduce the third derivative of the one loop prepotential,
$\partial_T^3f$,
\begin{equation}
\partial_T^3 f=-i(U-\ubar)^2 D_T\partial_T \partial_{\ubar} {\cal I}\ ,
\label{f3}
\end{equation}
where the covariant derivative $D_T = \partial_T + {2\over T-\tbar}$.

Since $T,U$ belong to the coset $O(2,2)/O(2)^2$, the classical duality
group is in general a subgroup of $O(2,2)$ restricted into the rational
numbers and in the simplest case it is $O(2,2;{\bf Z})\sim
SL(2,{\bf Z})_T\otimes SL(2,{\bf Z})_U$. Under $SL(2,{\bf Z})_T$
transformations,
\begin{equation}
T\rightarrow T_g\equiv {aT+b \over cT+d}\ ,
\label{sl2z}
\end{equation}
with $a,b,c,d$ integers satisfying $ad-bc=1$, while the physical quantity
$\cal I$ is modular invariant. It then follows from eq.(\ref{f3}) that
$\partial_T^3 f$ is a modular function of
weight 4 in $T$ and $-2$ in $U$.

Integrating equation (\ref{f3}) one can determine $f$ up to a quadratic
polynomial in $T$ and $U$,
\begin{equation}
f(T,U)=\int^{(T,U)}_{(T_0,U_0)}\{
dT'Q(U,U')(T-T')^2\partial_{T'}^3f(T',U')+
(T\leftrightarrow U, T'\leftrightarrow U')\}\ ,
\label{intf}
\end{equation}
where $(T_0,U_0)$ is an arbitrary point, and $Q(x,x')$ is the second
order differential operator,
\begin{equation}
Q(x,x')=\frac{1}{2}(x-x')^2{\partial}^2_{x'}+
(x-x'){\partial}_{x'}+1 \ .
\label{diff}
\end{equation}
The path of integration should not cross any singularity of
$\partial_T^3 f$, while the result of the integral depends on the
homology class of such paths. Different choices of homology classes of
paths change $f$ by quadratic polynomials in $T,U$. Moreover under a
modular transformation (\ref{sl2z}), $f$ does not transform covariantly.
Using its integral representation (\ref{intf}), we see that
it has a weight $-2$ up to an addition of a quadratic polynomial ${\cal
P}^g$ in $T,U$,
\begin{equation}
f(T_g,U)=(cT+d)^{-2}[f(T,U)+{\cal P}^g(T,U)]\ .
\label{intft}
\end{equation}
The same transformation properties hold for the $U$ variable, as
well as for the $T\leftrightarrow U$ exchange. These transformations
should leave the physical metric (\ref{metric}), (\ref{Idef}) invariant.
Hence, one must have
\begin{equation}
{\rm Im}\{ (\partial_T-\frac{2}{T- \tbar})
(\partial_U-\frac{2}{U- \ubar}) {\cal P}^g(T,U) =0\ ,
\label{real}
\end{equation}
which is satisfied only if ${\cal P}^g(T,U)$ is a quadratic polynomial
with real coefficients. In fact, we will see below that this
ambiguity is related to the non-trivial quantum monodromies.

Modular invariance of the full effective action implies that the
dilaton should also transform. Imposing the requirement that duality
transformations should be compensated by K\"ahler transformations one
finds,
\begin{equation}
S \rightarrow  S + c\frac{f_U + {\cal P}^g_U}{cT+d} + \lambda_g\, ,
\label{Stransf}
\end{equation}
up to an arbitrary additive axionic shift, $\lambda_g$. It follows that
in the presence of one loop corrections one can define an invariant
dilaton $S_{\rm inv}$ \cite{dkll},
\begin{equation}
S_{\rm inv}\equiv S+{1\over 2}f_{TU}\ ,
\label{Sinv}
\end{equation}
which however is not a special coordinate of $N=2$ K\"ahler geometry.

{}From a direct string computation of the one-loop metric and using
eq.(\ref{f3}), one finds the following world-sheet integral
representation of
$\partial_T^3 f$,
\begin{equation}
\partial_T^3 f=4\pi^2 \frac{U-\ubar}{(T-\tbar)^2}
\int{d^2\tau}\, C(\bar{\tau})\,
\sum_{p_L,p_R}p_L\bar{p}_{R}^3\, e^{\pi i\tau\, |p_{L}|^2}\,
e^{-\pi i \bar{\tau}\, |p_{R}|^2}\, ,
\label{fTTT}
\end{equation}
where the integration extends over the fundamental domain of the modular
parameter $\tau\equiv\tau_1+i\tau_2$, and $C$ is a $T$-independent
modular function of weight $-2$ with a simple pole at infinity due to the
tachyon of the bosonic sector. The summation inside the integral extends
over the left- and right-moving momenta of ${\bf T^2}$,
\begin{eqnarray}
p_L &=& \frac{1}{\sqrt{2\,{\makebox Im}T\,{\makebox Im}U}}\,
(n_1+m_1\tbar+m_2\ubar+n_2\tbar\ubar)                 \label{pL}\\
p_R &=& \frac{1}{\sqrt{2\,{\makebox Im}T\,{\makebox Im}U}}\,
(n_1+m_1T+m_2\ubar+n_2T\ubar)                         \label{pR}
\end{eqnarray}
with $m_1$, $m_2$, $n_1$ and $n_2$ integer numbers. The r.h.s.\ of
eq.(\ref{fTTT}) is indeed an analytic function of $T$ and $U$, as can be
verified by taking derivatives with respect to $\tbar$ or $\ubar$.
Using eqs (\ref{pL}), (\ref{pR}), one can easily show that the
resulting expressions are total derivatives in $\tau$ and vanish upon
integration.

At the plane $T=U$ there are two additional massless gauge multiplets
which enhance the gauge symmetry to $SU(2)\otimes U(1)$. They correspond
to lattice momenta (\ref{pL}), (\ref{pR}) with $n_1=n_2=0$ and
$m_2=-m_1=\pm 1$, so that $p_L=0$ and $p_R=\pm i\sqrt{2}$. The gauge
group is further enhanced at the two special points $T=U=i$ and
$T=U=\rho$ giving rise to $SO(4)$ and $SU(3)$, respectively. It
follows that the one-loop metric (\ref{Idef}) has a logarithmic
singularity of the form ${\cal I}\sim {2\over\pi}\ln|T-U_g|$ for $T$
close to $U_g$ ($\ne i,\rho$) where $g$ is an $SL(2,{\bf Z})_U$ element
(\ref{sl2z}). As a result, the one-loop prepotential behaves as
\begin{equation}
f(T,U) \rightarrow -{i\over\pi}[(cU+d)T-(aU+b)]^2\ln(T-U_g)\, .
\label{limf}
\end{equation}

One can now use duality symmetry to determine $f$. As mentioned above
$\partial_T^3 f$ is a modular function of weight $4$ in $T$ and $-2$ in
$U$. Moreover from its integral representation (\ref{fTTT}), it has a
simple pole at $T=U$ (modulo $SL(2,{\bf Z})_U$) with residue $(-2i/\pi)$
in accordance with (\ref{limf}), and vanishes as $T\rightarrow i\infty$.
These properties fix $\partial_T^3 f$ uniquely to:
\begin{equation}
\partial_T^3 f=-{2i\over\pi}\frac{j_T(T)}{j(T)-j(U)}
\left\{\frac{j(U)}{j(T)}\right\}
\left\{\frac{j_T(T)}{j_U(U)}\right\}
\left\{\frac{j(U)-j(i)}{j(T)-j(i)}\right\}\ ,
\label{f3T}
\end{equation}
where $j(T)$ is the meromorphic function with a simple pole with residue
1 at infinity and a third order zero at $T=\rho$.

Along the lines discussed above one can study a model with rank 3. The
scalar components of the vector multiplets are the dilaton $S$ and a
modulus $T$ which belongs to the coset $O(2,1)/O(2)$. The classical
duality symmetry acting on $T$ is $O(2,1;{\bf Z})\equiv SL(2,{\bf Z})$.
The prepotential is given by:
\begin{equation}
F=\frac{1}{2}ST^2 + f(T)~~~,
\label{prep}
\end{equation}
where $f(T)$ is the one loop correction to the classical prepotential
$\frac{1}{2}ST^2$. At generic point in the $T$-moduli space, the gauge
group is abelian, namely $U(1)^3$ including the vector partner of the
dilaton and the graviphoton. However at $T=i$ (mod $SL(2,{\bf Z})$) two
extra vector multiplets become massless, giving rise to an enhanced
gauge group $U(1)^2 \times SU(2)$. Consequently the one-loop metric must
have a singularity of the form, $\ln |T-i|$ for $T$ close to $i$. This
in turn implies that $f$ must behave as $(T-i)^2 \ln (T-i)$.

As in the rank 4 case discussed above, one can construct the K\"ahler
potential starting from the prepotential $F$ and check that the
requirement that the $SL(2,{\bf Z})$ transformations of $T$ should
be K\"ahler transformations implies that $f(T)$ transforms with weight
$-4$, up to additive terms that are at most quartic in $T$.

Under $SL(2,{\bf Z})$ duality transformation the one-loop metric must
transform covariantly. Using  the expression for the metric in terms of
$f$, one can easily deduce that $i\partial_T^5 f$ is a meromorphic form
of weight $6$ with respect to $T$, with a third order pole at $T=i$ and
vanishing for $T\rightarrow\infty$. To fix completely $i\partial_T^5 f$
one uses the knowledge of the $SU(2)$ beta function coefficient and
also the fact that the monodromy group of $f$ must be embeddable in
$Sp(6,{\bf Z})$ as dictated by $N=2$ supergravity, as we will explain in
the following. The result is:
\begin{equation}
\partial_T^5 f =-\frac{1}{18 \pi i}
\left\{\frac{j_T(T)}{j(T)-j(i)}\right\}^3\left\{\frac{j(i)}{j(T)}
\right\}^2 \left\{5+13\frac{j(T)}{j(i)}\right\}\, ,
\label{difff1}
\end{equation}
Expression (\ref{difff1}) can be used to verify the heterotic-type II
duality conjecture of ref.\cite{N=2str} in the weak coupling limit, that
is the agreement with the prepotential of the type II string compactified
on the Calabi-Yau threefold $X_{12}(1,1,2,2,6)$. Indeed, after
identifying $S$ and $T$ with the two $(1,1)$ moduli of the type II side
$t_{1,2}$ as $S=2t_2$ and $T=t_1$, one can verify the agreement, in the
large $S$ limit, of the first few terms in the $q_1\equiv\exp2\pi it_1$
expansion of the two expressions.
\vglue 0.4cm
{\elevenbf\noindent 3. Monodromies of the one-loop prepotential}
\vglue 0.2cm

Let us now discuss the monodromy group acting on $f$ in the rank 4 case
\cite{afgnt}. At the classical level there is the usual action of the
modular group acting on $T$ and $U$ upper half planes, namely $PSL(2,{\bf
Z})_T\otimes PSL(2,{\bf Z})_U$. $PSL(2,{\bf Z})_T$ is generated by the
two elements,
\begin{equation}
g_1: ~T\rightarrow -1/T ~~~~~~~~~~~~~g_2: ~T\rightarrow -1/(T+1)\ ,
\label{A}
\end{equation}
and similarly $PSL(2,{\bf Z})_U$ by the corresponding elements
$g_1',g_2'$. These generators obey the $SL(2,{\bf Z})$ relations
\begin{equation}
(g_1)^2 = (g'_1)^2 = (g_2)^3 = (g'_2)^3 = 1\ ,
\label{grel}
\end{equation}
and the relations implied by the fact that the two $PSL(2,{\bf Z})$'s commute.
There is also an exchange symmetry generator, namely:
\begin{equation}
\sigma: ~T\leftrightarrow U,
\label{C}
\end{equation}
which satisfies $\sigma^2=1$. Moreover $\sigma$ relates the two
$PSL(2,{\bf Z})$'s via $g'_1=\sigma g_1\sigma$ and $g'_2=\sigma
g_2\sigma$. The above relations can be thought of as the relations
among the generators of the fundamental group of our classical moduli
space in the following way: topologically each $PSL(2,{\bf Z})$
fundamental domain is a two-sphere $S$ ($S'$) with $3$ distinguished
points $x_1$ ($x'_1$), $x_2$ ($x'_2$) and $x_3$ ($x'_3$), which can be
taken to be the images of $i$, $\rho$ and $\infty$ by the $j$-function.
Associated with these three points we have generators $g_i$ ($g'_i$) of
the fundamental group of orders $2$, $3$ and $\infty$ respectively,
subject to the conditions $g_3g_2g_1=1$ and $g'_3g'_2g'_1=1$. The total
space is then the product of the two spheres $S$ and $S'$ minus
$\{x_i\}\times S'$ and $S\times\{x'_i\}$, $i=1,2,3$, and the fundamental
group of the resulting $4$-dimensional space is the product of the
fundamental groups of the two punctured spheres. Including $\sigma$,
we have the additional relations $g'_i=\sigma g_i \sigma$ and
$\sigma^2=1$.

In the quantum case we have singularities at $T=U$ and consequently
we must remove the diagonal in the product of the two punctured spheres
and and the fundamental group of the resulting space is a braid group.
One can adapt the results of ref.\cite{Scott} to the present case, and
obtain the following relations:
\begin{eqnarray}
&&g_3 g_2 g_1=\sigma^2,~~~~~~(g_1)^2=(g_2)^3=1\nonumber\\
&&g'_i=\sigma^{-1}g_i\sigma\nonumber\\
&&g_1\sigma^{-1}g_2\sigma=\sigma^{-1}g_2\sigma g_1\nonumber\\
&&\sigma g_i\sigma^{-1}g_i=g_i\sigma^{-1}g_i\sigma.
\label{qrelations}
\end{eqnarray}
The full fundamental group is generated by three elements $\sigma$,
$g_1$, $g_2$ subject to the above relations. In the quantum case
$\sigma^2$ corresponds  to moving a point around $T=U$ singularity, and
it will not be equal to the identity. Notice that if one sets
$\sigma^2=1$ one gets back the classical relations for the two commuting
$PSL(2,{\bf Z})$'s. However in the quantum case $\sigma^2\neq 1$ and the
two $PSL(2,{\bf Z})$'s do not commute anymore. Under $\sigma^2$, $f$
transforms as following:
\begin{equation}
Z_1 \equiv \sigma^2 : f(T,U) \rightarrow f(T,U) +2 (T-U)^2
\label{Z1}
\end{equation}
One can explicitly check the non commutativity
of $T$ and $U$ duality transformations using the integral
representation for $f$ given in (\ref{intf}).


Having the generators and relations of the fundamental group,
we will now determine the monodromy transformations of the
prepotential $f$. As explained in the previous section, under the
generators $g_1$, $g_2$ and $\sigma$, $f$ transforms according to
eq.(\ref{intft}) with three corresponding polynomials ${\cal P}^{g_1}$,
${\cal P}^{g_2}$ and ${\cal P}^{\sigma}$, quadratic in $T,U$ with real
coefficients. Imposing the group relations (\ref{qrelations}) and
eq.(\ref{Z1}), one can fix these polynomials in terms of 9 parameters
which correspond to the freedom of adding to $f$ a quadratic polynomial
in $T,U$ with real coefficients leaving the K\"ahler potential (\ref{Y})
invariant. In a particular base choice, one finds:
\begin{eqnarray}
{\cal P}^{g_1}&=&0\qquad\quad {\cal P}^{g_2}=2(T^2-1)\nonumber\\
{\cal P}^{\sigma}&=&(T-U)^2+(T-U)(-2UT+T+U+2)\ .
\label{poly}
\end{eqnarray}

The full monodromy group $G$ contains a normal abelian subgroup $H$,
which is generated by elements $Z_g$ obtained by conjugating $Z_1$ by an
element $g$ which can be any word in the $g_i$'s, $g'_i$'s and their
inverses. $Z_g$ corresponds to moving a point around the singularity
$T=U_g$, where the prepotential behaves as shown in (\ref{limf}). A
general element of $H$ is obtained by a sequence of such transformations
and shifts $f$ by:
\begin{equation}
f\rightarrow f +2\sum_i N_i ((c_iU+d_i)T-(a_iU+b_i))^2
\equiv f +\sum_{n,m =0}^2 c_{nm} T^n U^m ~~~ N_i \in {\bf Z}
\label{He}
\end{equation}
with $a_i,b_i,c_i,d_i$ corresponding to some $SL(2,{\bf Z})$ elements for
each $i$. Since the polynomial entering in (\ref{He}) has 9 independent
parameters $c_{nm}$, it follows that $H$ is isomorphic to $Z^9$. The set
of all conjugations of $H$ by elements generated by $g_i$'s and $g'_i$'s
defines a group of (outer) automorphisms of $H$ which is isomorphic to
$PSL(2,{\bf Z})\times PSL(2,{\bf Z})$, under which $c_{nm}$ transform as
$(3,3)$ representation (in this notation the two $PSL(2,{\bf Z})$'s act
on the index $n,m$ respectively). Moreover, the conjugation by $\sigma$
defines an automorphism which interchanges the indices $n$ and $m$ in
$c_{nm}$. Thus the set of all conjugations of $H$ is isomorphic to
$O(2,2;{\bf Z})$, under which the $c_{nm}$'s transform as a second rank
traceless symmetric tensor. Finally, the quotient group $G/H$ is
isomorphic to $O(2,2;{\bf Z})$, therefore $G$ is a group involving 15
integer parameters. On the other hand, $G$ is not the semidirect product
of $O(2,2;{\bf Z})$ and $H$, since $O(2,2;{\bf Z})$ is not a subgroup of
$G$, as it follows from the quantum relations (\ref{qrelations}).
Of course for physical on-shell quantities the group $H$ acts trivially
and therefore one recovers the usual action of $O(2,2;{\bf Z})$.

In addition to the above monodromies there is also the axionic shift
$D$, defined in eq.(\ref{axionshift}), and the full perturbative group of
monodromies is the direct product of $G$ with this abelian translation
group.

The above monodromy group structure is best exploited in a field basis
where all monodromies act linearly \cite{ceres,afgnt}. To this end
we use the formalism of the standard $N{=}2$ supergravity
where the physical scalar fields $Z^I$ of vector multiplets are
expressed as
$Z^I=X^I/X^0$, in terms of the constrained fields $X^I$ and $X^0$. This
is a way to include the extra $U(1)$ gauge boson associated with the
graviphoton which has no physical scalar counterpart. In our case we have
\begin{equation}
S={X^s\over X^0} ~~~ T={X^2\over X^0} ~~~ U={X^3\over X^0}
\label{xbasis}
\end{equation}
and the prepotential (\ref{F}) is a homogeneous polynomial of
degree 2, $F(X^I)=(X^0)^2 F(S,T,U)$. The K\"ahler potential $K$ is
\begin{equation}
K=-\log i(\bar{X}^I F_I-X^I \bar{F}_I)\ ,
\label{Kxbasis}
\end{equation}
where $F_I$ is the derivative of $F$ with respect to $X^I$ and
$I=0,s,2,3$. This has a generalization in basis where $F_I$ is not the
derivative of a function $F$ \cite{ceres}.

Clearly the symplectic transformations acting on $(X^I,F_I)$ leave the
K\"ahler potential invariant. Since the monodromy group leaves $K$
invariant, we expect it to be a subgroup of the symplectic group
$Sp(8)$. A general symplectic transformation is
\begin{equation}
\pmatrix{X^I\cr F_I\cr}\rightarrow \pmatrix{{\bf a}&{\bf b}\cr
{\bf c}&{\bf d} \cr}
\pmatrix{X^I\cr F_I\cr}
\label{symp}
\end{equation}
where ${\bf a}$, ${\bf b}$, ${\bf c}$, ${\bf d}$ are $4\times 4$ matrices
satisfying the defining relations of the symplectic group, ${\bf a}^t
{\bf c}- {\bf c}^t{\bf a}=0$, ${\bf b}^t{\bf d}-{\bf d}^t{\bf b}=0$ and
${\bf a}^t {\bf d}-{\bf c}^t{\bf b}={\bf 1}$. Under this transformation,
the vector kinetic term Im${\cal F}^I_{\mu\nu} {\overline N}_{IJ} {{\cal
F}^J}^{\mu\nu}$ transforms as $N \rightarrow ({\bf c}+{\bf d}N)({\bf
a}+{\bf b}N)^{-1}$. It follows that for ${\bf b}\ne 0$ the gauge
coupling gets inverted and therefore in a suitable basis the
perturbative transformations must have ${\bf b}=0$. When ${\bf b}=0$ the
symplectic constraints become ${\bf d}^t={\bf a}^{-1}$ and ${\bf c}=
{{\bf a}^t}^{-1}{\tilde{\bf c}}$ with ${\tilde{\bf c}}$ an arbitrary
symmetric matrix. In this case, the vector kinetic term changes by
${\tilde{\bf c}}_{IJ}\makebox{Im}{\cal F}^I{\cal F}^J$ which, being a
total derivative, is irrelevant at the perturbative level. However at
the non-perturbative level, due to the presence of monopoles, the matrix
${\tilde{\bf c}}$ must have integer entries.

At the classical level one can easily verify that the $PSL(2,{\bf Z})_T$
transformation (\ref{sl2z}) (and similarly $PSL(2,{\bf Z})_U$
transformation given by interchanging $X^2$ with $X^3$ and $F_2$ with
$F_3$) acts on $X^I$ and $F_I$  by a symplectic matrix, whose entry ${\bf
b}$ is however different from zero. It is therefore convenient to
perform a change of basis into $(X^I,F_I)$, where $I=0,1,2,3$ and
$X^1=F_s$ and $F_1=-X^s$. In the new basis the tree level $O(2,2;{\bf
Z})$ transformations are block diagonal, i.e.\ ${\bf b}={\bf c}=0$ and
${\bf d}={{\bf a}^t}^{-1}$.

Having found a basis which is appropriate for the perturbative
monodromies one can proceed to recast the previous discussion about the
monodromy transformations of $f$ in a linear, symplectic form and get
the symplectic matrices corresponding to the braid group generators. We
will not give their explicit form here, but just mention their basic
features. They have ${\bf b}=0$, that is they are of the form:
\begin{equation}
\pmatrix{{\bf a}&{0}\cr {{\bf a}^t}^{-1}{\tilde{\bf c}}&{{\bf a}^t}^{-1}
\cr}\ .
\label{symp2}
\end{equation}
The matrices $\tilde{\bf c}$ turn out to be symmetric and satisfy
Tr$\eta{\tilde{\bf c}}=0$, where $\eta$ is the $O(2,2)$ metric $\eta={\rm
diag} (\sigma_1,-\sigma_1)$.

The abelian group $H$ introduced in (\ref{He}) is generated by symplectic
matrices (\ref{symp2}) with ${\bf a}={\bf 1}$, and
\begin{equation}
{\tilde{\bf c}}=\sum_i 2N_ig_i^t\pmatrix{\sigma_1&0\cr 0&
\sigma_1-2\cr}g_i\ ,
\label{newH}
\end{equation}
where $g_i$ can be chosen for instance as $PSL(2,{\bf Z})_T$ matrices.
These matrices are traceless with respect to $\eta$, and depend on 9
integer parameters. They form the 9-dimensional representation of
$O(2,2;{\bf Z})$ corresponding to the second rank symmetric traceless
tensors, as explained previously.

Finally the axionic shift in the above symplectic basis corresponds to
${\bf a}=1$ and ${\bf c}=-\lambda \eta$, which commutes with the above
matrices of $G$, as expected. The parameter $\lambda$ should also be
quantized at the non-perturbative level. In this way one generates all
possible symmetric $4\times 4$ lower off-diagonal matrices depending on
10 integer parameters, the trace part being generated by $\eta$. The full
monodromy group is generated by 4 generators: $g_1$, $g_2$, $\sigma$ and
the axionic shift.
\vglue 0.4cm
{\elevenbf\noindent 4. Subgroups of Rigid Monodromies}
\vglue 0.2cm

At the semiclassical level, one $O(2,2)$ model embeds four distinct
gauge groups: $U(1)\otimes U(1)$ at a generic point of the moduli space,
$SU(2)\otimes U(1)$ along the $T=U$ line, $SO(4)$ at $T=U=i$, and
$SU(3)$ at $T=U=\rho$. We will discuss now the relation between
the monodromies of the respective $N{=}2$ globally supersymmetric
(rigid) Yang-Mills theories and the superstring monodromy group.

In the case of gauge group of rank 2, the rigid monodromy group
is a subgroup of $Sp(4,{\bf Z})$. It is generated by Weyl reflections
\begin{equation}
{\bf a}_k~=~{1\!\! 1}-2\frac{{\bf\alpha}_k
\otimes{\bf\alpha}_k}{{\bf\alpha}_{k}^2}\label{arig}\ ,\end{equation}
where ${\bf \alpha}_k,~k=1,2$ are simple roots of the gauge group.
The corresponding monodromy matrices can be written as
\begin{equation} M_k ~=~
\pmatrix{{\bf a}_k &0\vspace{1mm}\cr
{{\bf a}_k^t}^{-1}\tilde{\bf c}_k& {{\bf a}_k^t}^{-1}},~~~~~~
\tilde{\bf c}_k= {\bf\alpha}_k\otimes{\bf\alpha}_k\ .
\label{mrig}
\end{equation}

In the field basis in which the classical duality $O(2,2;{\bf Z})$
group is realized as block diagonal $Sp(8,{\bf Z})$ matrices discussed
in the previous section, the surfaces of enhanced symmetries in the
$\Gamma_{2,2}$ lattice (\ref{pL}-\ref{pR}), $p_L=0$ and $p_R^2=2$, are
defined by the equation
\begin{equation}
p_L\sim\frac{n_{I}^{{\bf\alpha}_i}X^I}{X^0}=n_1^{{\bf\alpha}_i}
+n_2^{{\bf\alpha}_i}TU+ m_1^{{\bf\alpha}_i}T+m_2^{{\bf\alpha}_i}U=0\ ,
\label{surf}
\end{equation}
for the particular choice of $n_I^{{\bf\alpha}_i}=
(n_1,n_2,m_1,m_2)^{{\bf \alpha}_i} \in\Gamma_{2,2}$ obeying
$n_I\,n^I=2(n_1n_2-m_1m_2)=2$. These $\Gamma_{2,2}$ vectors represent the
root vectors ${\bf\alpha}_i$ of enhanced symmetries corresponding to
gauge multiplets that become massless on the surface. It is well known
that Weyl reflections associated with these roots  can be represented as
$O(2,2;{\bf Z})$ duality transformations \cite{du}; they correspond to
Weyl reflections $({\bf a}^{{\bf\alpha}_i})^I_{~J}~=
\delta^I_J-{n^{{\bf\alpha}_i}}^I{n^{{\bf\alpha}_i}}_J$
in the surfaces (\ref{surf}). In the Table below, we give a list of
duality transformations and the respective $O(2,2)$
matrices,\footnote{As usual, these matrices are defined up to an overall
multiplication by $-{1\!\! 1}$.} for Weyl reflections associated with
the simple roots.
\begin{center}
\bf $\bf O(2,2)$ embeddings of Weyl reflections
\end{center}
\begin{tabular}[b]{|c|c|@{}c@{}c@{}} \cline{1-2} & & & \\
& $n_I^{{\bf\alpha}_1}=(0,0,1, -1)$ &  &\\[4mm]
$T=U$ & $\sigma :~~~T\rightarrow U$ & &\\[4mm] $SU(2)$ &
${\bf a}^{{\bf\alpha}_1}={\scriptstyle \pmatrix{1&0&0&0\cr
0&1&0&0\cr 0&0&0&1\cr 0&0&1&0}}$ & &\\ & & & \\ \cline{1-3}
& & \hfill\vline  &\\ & &
{}~~~~~~~~$n_I^{{\bf\alpha}_2}=(1,1,0,0)$ \hfill\vline &\\[4mm]
$T=U=i$ & same as
above & ~~~~~$g_1\sigma g_1^{-1} :~~T\rightarrow -1/U$\hfill\vline&\\[4mm]
$SO(4)$ & &~~${\bf a}^{{\bf\alpha}_2}={\scriptstyle \pmatrix{0&-1&0&0\cr
-1&0&0&0\cr 0&0&1&0\cr 0&0&0&1}}$ \hfill\vline&\\ & &\hfill\vline & \\
\hline & & \hfill\vline&\hfill\vline\\
& & ~~~~~~~~$n_I^{{\bf\alpha}_3}=(1,1,0,1)$ \hfill\vline &
{}~~~~~~~~$n_I^{{\bf\alpha}_4}=(1,1,1,0)$ \hfill\vline\\[4mm]
$T=U=\rho$& same as above
& ~$g_2^{-1}\sigma g_2 :~~T\rightarrow -(U+1)/U$~\hfill\vline &
{}~$g_2\sigma g_2^{-1}:~~ T\rightarrow -1/(U+1)$
\hfill\vline\\[4mm] $SU(3)$& &
{}~~${\bf a}^{{\bf\alpha}_3}={\scriptstyle \pmatrix{0&-1&0&-1\cr
-1&0&0&-1\cr 1&1&1&1\cr 0&0&0&1}}$\hfill\vline&
{}~~${\bf a}^{{\bf\alpha}_4}={\scriptscriptstyle \pmatrix{0&-1&-1&0\cr
-1&0&-1&0\cr 0&0&1&0\cr 1&1&1&1}}$~\hfill\vline\\
& & \hfill\vline&\hfill\vline  \\ \hline
\end{tabular}
\vskip 5mm

The appearance of massless states  at the enhanced symmetry surfaces
gives rise to the logarithmic singularities  which modify the classical
monodromies. Near the surface (\ref{surf}), the singular part of $f_I$ is
\begin{equation}
f_I=-\frac{2i}{\pi}n_I^{\alpha_i}{n^{\alpha_i}}_J X^J\log(
\frac{n_K^{\alpha_i}X^K}{X^0})\ .
\end{equation}
The identification of Weyl reflections as duality transformations is very
useful if one wants to study the embedding of rigid monodromies $M_k\in
Sp(4,{\bf Z})$ of eq.(\ref{mrig}) into $Sp(8,{\bf Z})$, as a subgroup
of the superstring monodromy group \cite{Lust}. The explicit form of the
respective matrices depends however on the  choice of the base point
which amounts to the freedom of adding to $f$ a polynomial quadratic in
$T$ and $U$ with real coefficients or equivalently, of conjugating the
monodromy matrices by the elements of the normal abelian subgroup $H$.
One can ask the question whether a base exists such that the monodromies
associated with Weyl reflections in singular surfaces have the form of
rigid monodromies of eq.(\ref{mrig}). The answer is affirmative. Indeed
one can change
the base from the one used in the previous section and in
ref.\cite{afgnt} to the new one which corresponds to
\begin{eqnarray}
f &\rightarrow& f+1+2T+T^2-2TU-2TU^2-4T^2U+\epsilon (TU -1)^2\  ,\\
S &\rightarrow& S+2U+4T-2\epsilon TU\ , \nonumber
\end{eqnarray}
where $\epsilon$ is still a free parameter. It is easy to verify that with
an additional dilaton shift accompanying $\sigma$ transformation,
 $\lambda_{\sigma}=1$, the $Sp(8,{\bf Z})$ monodromies
associated with Weyl reflections
$\sigma$, $g_1\sigma g_1^{-1}$ and $g_2\sigma g_2^{-1}$
become precisely of the form of
eq.(\ref{mrig}), with the $O(2,2)$ matrices $\bf a$ as given in the
Table and
\begin{equation}
(\tilde{{\bf c}}_k)_{IJ}=2 n_I^{{\bf\alpha}_k}{n^{{\bf\alpha}_k}}_J\ .
\end{equation}
Notice that the third $SU(3)$ Weyl reflection $g_2^{-1}\sigma   g_2$ is
not of the above form, similarly to the rigid case, where only the Weyl
reflections corresponding to the simple roots are of this form \cite{klt2}.
Finally the three Weyl reflections are related by conjugation with the
Coxeter element $g_2$.
Hence the rigid perturbative monodromies associated with all possible
enhanced
gauge symmetries are glued together in superstring theory, embedded in one
monodromy group in a very natural way. {}From the above discussion it is
clear that this conclusion applies also to  arbitrary (4,4)
compactifications.

Let us now turn to the question of non-perturbative monodromies. As in the
rigid case, we expect that in the quantum moduli space, the
singularities corresponding to enhanced gauge symmetries disappear and
are replaced by the ones associated to monopole or dyonic states that
become massless. If we label the dyonic charge by $(q_I;g^J)$ where $q$
and $g$ are the electric and magnetic charges respectively, then the
monodromy due to this state becoming massless is given by the following
symplectic matrix:
\begin{equation} M_{(q,g)} ~=~
\pmatrix{{{1\!\! 1}- g\otimes q} &{-\frac{1}{2}g\otimes g}\vspace{1mm}\cr
{2 q\otimes q}& {{1\!\! 1}+q\otimes g}}.
\label{mrig1}
\end{equation}
The factors of $2$ in the above equation are due to our normalization of the
charge vectors.
The three singularities
associated with the Weyl reflections $\sigma$,
$g_1 \sigma g_1^{-1}$ and $g_2\sigma g_2^{-1}$ each split into two
singularities;
one corresponding to a pure monopole state and the second that to a dyonic
state. The pure monopole state carries the magnetic charge $g^I = \eta^{IJ}
n_J^{\alpha_k}$ with $n_J^{\alpha_k}$ defined in the Table. The monodromy
matrix associated to this monopole is then obtained by substituting the
magnetic charge in (\ref{mrig1}). The monodromy corresponding to the
dyonic state can be obtained by using the fact that the product of the
monodromies
associated to the pure monopole and the dyonic state equals the
perturbative monodromy corresponding to $\alpha_k$. More explicitly
the dyonic charges $(q,g)$ are given by $q_I=n_I^{\alpha_k}$ and
$g^I = -\eta^{IJ} n_J^{\alpha_k}$. Finally the
monodromies
for the two dyonic states associated with the third $SU(3)$ root
$\alpha_3$ are obtained by
the action of the Coxeter element $g_2$.

\vglue 0.6 cm {\elevenbf\noindent Acknowledgements} \vglue 0.2 cm

This work was supported in part by the National Science Foundation under
grant PHY--93--06906, in part by the EEC contracts SC1--CT92--0792,
CHRX-CT93-0340 and SC1*CT92--0789, in part by the U.S. DOE grant
DE-FG03-91ER40662 Task C, and in part by CNRS--NSF grant INT--92--16146.


\begin{thebibliography}{99}
\bibitem{centext}  For recent reviews,  see:  A. Ceresole, R. D'Auria
and S. Ferrara, hep-th/9509160; M.J. Duff, hep-th/9509106;
J.H. Schwarz, hep-th/9509148.
\bibitem{N=4} A. Sen, \pl 329 (1994) 217; C. Vafa and E. Witten, \np 431
(1994) 3; for a review see D. Olive, hep-th/9508089.
\bibitem{N=2} N. Seiberg and E. Witten, \np 431  (1994)  484;  A. Klemm, W.
Lerche, S. Yankielowicz and S. Theisen, \pl 344  (1995)  169; P.C. Argyres and
A. Faraggi, Phys.\  Rev.\  Lett.\  74  (1995)  3931.
\bibitem{strstr} C.M. Hull and P. Townsend, \np 438 (1995) 109; M.J.
Duff, \np 442 (1995) 47; E. Witten, \np 443 (1995) 85.
\bibitem{ceres} A. Ceresole, R. D'Auria, S. Ferrara and A. Van Proeyen,
\np 444 (1995) 92.
\bibitem{N=2str} S. Kachru and C. Vafa, hep-th/9505105; S. Ferrara, J.A.
Harvey, A. Strominger and C. Vafa, hep-th/9505162.
\bibitem{du} For a review, see A. Giveon, M. Porrati and E. Rabinovici,
Phys.\ Rep.\ 244 (1994) 77.
\bibitem{afgnt} I. Antoniadis, S. Ferrara, E. Gava, K.S. Narain and T.R.
Taylor, \np 447 (1995) 35.
\bibitem{dkll} B. de Wit, V. Kaplunovsky, J. Louis and D. L\"ust,
\np 451 (1995) 53.
\bibitem{CY} P. Candelas, X. De la Ossa, P. Green and L. Parkes, \np
359 (1991) 21; P. Candelas, X. De la Ossa, A. Font, S. Katz and
D. Morrison, \np 416 (1994) 481; S. Hosono, A. Klemm, S. Theisen and
S.T. Yau, Comm. Math. Phys. 167 (1995) 301.
\bibitem{candc} P. Candelas, P. Green and T. H\"ubsch, \np 330 (1990) 49;
P. Candelas and X. De la Ossa, \np 342 (1990) 246.
\bibitem{andy} A. Strominger, hep-th/9504090.
\bibitem{cheks} V. Kaplunovsky, J. Louis and S. Theisen,
hep-th/9506110; A. Klemm, W. Lerche and P. Mayr, \pl 357 (1995) 313; I.
Antoniadis, E. Gava, K.S. Narain and T.R. Taylor, as reported by K.S.
Narain at the {\it Trieste Conference on Mirror Symmetry and S-Duality},
5-9 June 1995.
\bibitem{monodr} M. Bill\'o, A. Ceresole, R. D'Auria, S. Ferrara, P. Fr\`e,
T. Regge, P. Soriani and A. Van Proeyen, hep-th/9506075; S. Kachru, A.
Klemm, W. Lerche, P. Mayr and C. Vafa, hep-th/9508155; I. Antoniadis and
H. Partouche, hep-th/9509009.
\bibitem{w2g} I. Antoniadis, E. Gava, K.S. Narain and T.R. Taylor,
hep-th/9507115.
\bibitem{yuka} I. Antoniadis, E. Gava, K.S. Narain and T.R. Taylor,
\np 407 (1993) 706.
\bibitem{Scott} G.P. Scott, Proc. Camb. Phil. Soc. 68 (1970) 605.
\bibitem{Lust} G.L. Cardoso, D. L\"{u}st and T. Mohaupt, hep-th/9507113.
\bibitem{klt2} A. Klemm, W. Lerche and S. Theisen, hep-th/9505150.
\end{thebibliography}
\end{document}